\documentclass[10pt,aps,prb,reprint,superscriptaddress,amsfonts,amssymb,amsmath,floatfix,
longbibliography]{revtex4-2}
\usepackage{microtype}
\usepackage{lmodern}
\usepackage[utf8]{inputenc}
\usepackage[T1]{fontenc}
\usepackage{graphicx}
\usepackage{hyperref}
\hypersetup{
    colorlinks=true,
    linkcolor=blue,
    citecolor=blue,   
    urlcolor=blue
}
\usepackage{epstopdf}%
\usepackage{xcolor}
\newcommand{\change}[1]{\textcolor{black}{#1}}

\begin{document}

\title{Pb-intercalated epitaxial graphene on SiC: Full insight into band structure and orbital character of interlayer Pb, and charge transfer into graphene}

\author{Bharti Matta}
\email{b.matta@fkf.mpg.de}
\affiliation{Max-Planck-Institut für Festkörperforschung, Heisenbergstraße 1, 70569 Stuttgart, Germany}
\author{Philipp Rosenzweig}
\affiliation{Max-Planck-Institut für Festkörperforschung, Heisenbergstraße 1, 70569 Stuttgart, Germany}
\affiliation{Physikalisches Institut, Universität Stuttgart, Pfaffenwaldring 57, 70569 Stuttgart, Germany}
\author{Kathrin K{\"u}ster}
\affiliation{Max-Planck-Institut für Festkörperforschung, Heisenbergstraße 1, 70569 Stuttgart, Germany}
\author{Craig Polley}
\affiliation{MAX IV Laboratory, Lund University, Fotongatan 2, Lund 22484, Sweden}
\author{Ulrich Starke}
\affiliation{Max-Planck-Institut für Festkörperforschung, Heisenbergstraße 1, 70569 Stuttgart, Germany}

\begin{abstract}
Intercalation is a robust approach for modulating the properties of epitaxial graphene on SiC and stabilizing two-dimensional (2D) intercalant layers at the graphene/SiC interface. In this work, we present synchrotron-based angle resolved photoelectron spectroscopy (ARPES) measurements focussing on the band structure of intercalated Pb under a single layer of epitaxial graphene. The interlayer Pb exhibits a metallic character, a {\mbox{(1${\times}$1)}} registry with respect to SiC, and free electron-like bands to a first order. Divergences from the free electron approximation include various band splittings and gaps throughout the Pb Brillouin zone. Light polarization dependent ARPES measurements indicate a predominant out-of-plane orbital character for the Pb bands, suggesting potential interactions between the interlayer Pb and graphene's $\pi$ orbitals  that may induce proximity effects in graphene. Density functional theory (DFT) calculations for a {\mbox{(1${\times}$1)}} Pb monolayer on SiC show a reasonable qualitative agreement with the experimentally observed interlayer bands as well as the polarization dependent measurements. Finally, temperature dependent ARPES measurements reveal that the nearly charge-neutral graphene layer involves charge transfer from both the interlayer Pb and the substrate SiC. 
\end{abstract}

\maketitle

\section{INTRODUCTION}
Epitaxial graphene on SiC has emerged as a prominent platform for intercalating different elements and modifying graphene's electronic structure \cite{RIEDL2009,MCCHESNEY2010,EMTSEV2011,MARCHENKO2016,FORTI2016,LINK2019}, with the advantage of wafer-scale graphene growth on a technologically adaptable substrate \cite{BERGER2004,EMTSEV2009,YU2011,KRUSKOPF2016}. Over the past few years, the focus has gradually been shifting from the modification of graphene's properties to understanding the quantum-confined intercalant itself. The intercalant layers can form well-ordered structures at the graphene/SiC interface and present their own unique electronic band characteristics \cite{HAYASHI2017,FORTI2020,ROSENZWEIG2020,BRIGGS2020,SOHN2021,MATTA2022,SCHMITT2024}. 
\\Pb has raised particular interest as an intercalant for epitaxial graphene on SiC \cite{YURTSEVER2016,CHEN2020,YANG2021,GRUSCHWITZ2021,HU2021,MATTA2022,GHOSAL2022,HAN2023,otherHAN2023,SCHADLICH2023,VERA2024,SCHOLZEL2024,WANG2021,BROZZESI2024,HAN2024,WANG2024,GRUSCHWITZ2024}, with the motivation of having graphene in proximity to 2D Pb, which might induce superconductivity and spin-orbit coupling \change{(SOC)} effects in graphene \cite{KLIMOVSKIKH2017,OTROKOV2018,CHERKEZ2018,PASCHKE2020}. Pb-intercalated quasi-freestanding monolayer graphene (Pb-QFMLG) on SiC exhibits near charge-neutrality, the underlying reason for which is under active discussion \cite{MATTA2022,SCHADLICH2023,SCHOLZEL2024}. A systematic investigation is required to disentangle the interplay of different charge transfer contributions from both the intercalated Pb and the SiC substrate to the graphene. To date, most studies on Pb-intercalated epitaxial graphene have concentrated on the atomic arrangement of the intercalated Pb layer \cite{YURTSEVER2016,YANG2021,GRUSCHWITZ2021,HU2021,MATTA2022,GHOSAL2022,HAN2023,otherHAN2023,SCHADLICH2023,VERA2024,SCHOLZEL2024,HAN2024,WANG2024}. 
A number of studies discuss aspects of the electronic band structure of interlayer Pb
 \cite{MATTA2022,SCHADLICH2023,VERA2024,SCHOLZEL2024,WANG2021,BROZZESI2024}, yet a comprehensive study of the Pb bands and their properties remains absent.
\\The current work discusses in depth the electronic band structure of interlayer Pb using synchrotron-based angle resolved photoelectron spectroscopy (ARPES) at different photon energies and light polarizations, revealing different band features, their atomic orbital character as well as the 2D nature of interlayer Pb. The experimental data are compared with a free electron approximation and density functional theory (DFT) calculations for a {\mbox{(1${\times}$1)}} Pb monolayer on SiC. Furthermore, we discuss the charge transfer involved in rendering the Pb-QFMLG nearly charge-neutral.

\section{EXPERIMENT AND THEORY}\label{sec2}

\begin{figure*}[t]
	\centering
    \includegraphics{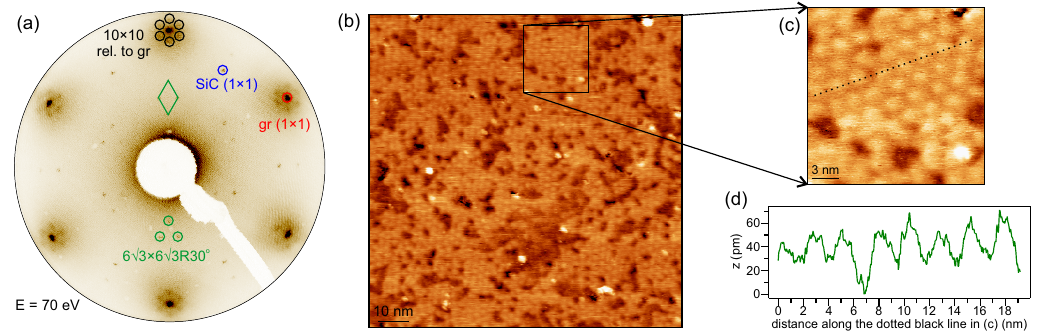}
	\caption{(a) LEED pattern of Pb-QFMLG at $70$ eV. (b) STM image obtained across a sample area of $100\times100~\mathrm{nm}^2$, $U_\mathrm{bias}$ = $-400$ mV, $I$ = $300$ pA. (c) A zoomed in measurement on the black box marked in (b); $20\times20~\mathrm{nm}^2$, $U_\mathrm{bias}$ = $-400$ mV, $I$ = $400$ pA. (d) Height profile along the black dotted line in (c) showing an average periodicity of $\approx 2.5~\mathrm{nm}$, reflecting a {\mbox{10${\times}$10}} periodicity with respect to graphene ($a = 2.46~{\AA}$).}
	\label{Fig_1}
\end{figure*}

\subsection{Sample preparation and characterization}
Single crystalline, on-axis, $n$-doped 6H-SiC(0001) wafer pieces (SiCrystal GmbH) served as substrates for graphene growth, performed \emph{ex situ} in a graphite susceptor within an inductively heated reactor. The substrates were initially etched using molecular hydrogen at $800$ mbar and $1550$~$^{\circ}\mathrm{C}$, resulting in the removal of any polishing scratches, yielding atomically flat terraces \cite{RAMACHANDRAN1998,SOUBATCH2005}. Afterwards, the substrates were graphitized by annealing to $1460$~$^{\circ}\mathrm{C}$ in an argon atmosphere at $800$ mbar, for $4$ to $5$ min, leading to the development of several $\mu$m wide terraces covered by the $(6\sqrt{3}\times6\sqrt{3})\mathrm{R}30^\circ$ reconstruction of the carbon buffer layer on SiC, also commonly referred to as zerolayer graphene (ZLG) \cite{RIEDL2010}. 
\\The ZLG samples were transferred into an ultrahigh vacuum (UHV) chamber and outgassed at $700$~$^{\circ}\mathrm{C}$ for about $30$ min. For Pb-intercalation, Pb was deposited on ZLG at room temperature, at a rate of about 4~{\AA}/min for $10$ to $20$ min from a Knudsen cell. This was followed by sequential annealing of the sample from $200$~$^{\circ}\mathrm{C}$ up to $550$~$^{\circ}\mathrm{C}$, with each annealing step lasting for $20$ to $60$ min \cite{MATTA2022}. Intercalation was monitored using low energy electron diffraction (LEED) and further annealing of the sample at higher temperatures revealed signs of deintercalation. Pb-deposition and annealing cycles were repeated once or twice to achieve homogeneous intercalation. 
\\The Pb-QFMLG samples were carried to the MAX IV synchrotron facility (Lund, Sweden) in a Ferrovac UHV suitcase with a base pressure below $5\times10^{-10}$ mbar. ARPES measurements were carried out at the Bloch beamline equipped with a DA30-L hemispherical analyzer from ScientaOmicron. Its electronic deflection mode allows mapping the entire surface Brillouin zones (BZs) of Pb and graphene in a single measurement, without the need of repositioning the sample. All the measurements were performed at a sample temperature of $\approx 18$ K with linear horizontal light polarization, unless stated otherwise. The maximal and minimal energy resolutions (considering the spectrometer and the beamline) used were about $15$ meV and $64$ meV, respectively. Room temperature scanning tunneling microscopy (STM) measurements were also carried out at the Bloch beamline using an Omicron VT-XA STM with electron-beam annealed tungsten tips. The bias voltage was applied to the tip relative to the sample. Different samples were used in this work, yielding consistent results and thereby confirming the reproducibility of our intercalation procedure.

\subsection{DFT calculations}
DFT calculations for a {\mbox{(1${\times}$1)}} monolayer Pb on SiC were performed using the Perdew–Burke–Ernzerhof generalized gradient approximation exchange-correlation functional, as implemented in the Quantum ESPRESSO package \cite{GIANNOZZI2009,GIANNOZZI2017}. SOC was included through the use of fully relativistic projector augmented wave pseudopotentials. A six bilayer SiC slab model was used, with one side having Pb atoms attached in the top-site position to the surface Si atoms and the other side having hydrogen atoms attached in the top-site position to the surface C atoms, with $20~\AA$ of vacuum separation between the slabs. The plane-wave cutoff energy and the $k$-point mesh sampling density were 70~Ry and 16$\times$16$\times$1, respectively. The in-plane lattice parameter for Pb was fixed to $\approx 3.08~{\AA}$, corresponding to the experimentally measured lattice parameter for bulk 6H-SiC, while all atoms in the slab were relaxed along the $c$-axis to a force convergence threshold of $5\times{10^{-7}}$ Ry/bohr. The resulting distance between the Pb atoms and the surface Si atoms was $\approx 2.8~{\AA}$, which is in close agreement with X-ray standing waves (XSW) measurements for a Pb-QFMLG \cite{SCHADLICH2023}.

\section{RESULTS AND DISCUSSION} 

\subsection{Basic atomic structure characterization of Pb-QFMLG}

The so-called ZLG on SiC is partially covalently bonded to the Si dangling bonds and therefore, does not exhibit the characteristic electronic properties of a freestanding monolayer graphene. In other words, ZLG does not show the well-known Dirac cone dispersion of graphene \cite{EMTSEV2008,RIEDL2009}.
Upon intercalation of ZLG, the intercalant atoms saturate or interact with the Si dangling bonds, lifting the ZLG and transforming it into quasi-freestanding monolayer graphene on SiC (QFMLG). This QFMLG exhibits the true electronic properties of graphene, and hence shows the Dirac cone dispersion.
\\After Pb intercalation, the diffraction spots corresponding to the $(6\sqrt{3}\times6\sqrt{3})\mathrm{R}30^\circ$ reconstruction of ZLG on SiC are largely suppressed in the LEED pattern [Fig.~\ref{Fig_1}(a)]. The first order diffraction spots of graphene are the most intense when compared to the first order SiC spots as well as the $(6\sqrt{3}\times6\sqrt{3})\mathrm{R}30^\circ$ reconstruction spots. Apart from that, additional diffraction spots are visible around the first order graphene spots, which correspond to an average periodicity of {\mbox{(10${\times}$10)}} relative to the graphene lattice vectors \cite{GRUSCHWITZ2021,MATTA2022,SCHADLICH2023,VERA2024}. Fig.~\ref{Fig_1}(b) shows an overview STM image for a Pb-QFMLG, which indicates that there is no $(6\sqrt{3}\times6\sqrt{3})\mathrm{R}30^\circ$ reconstruction inherent to a ZLG \footnote{The STM measurements shown in Fig.~\ref{Fig_1} do not present atomic resolution.}. A small-scale measurement in Fig.~\ref{Fig_1}(c) reveals a periodic corrugation, corresponding to a height elevation. A height ($z$) profile [see Fig.~\ref{Fig_1}(d)] along the black dotted line in Fig.~\ref{Fig_1}(c) gives an average distance of $\approx 2.5$ nm between the respective maxima, which is about $10$ times the lattice vector length of the graphene unit cell, consistent with the {\mbox{(10${\times}$10)}} superstructure spots observed in LEED. The weak long range order of this superstructure in STM explains the weakness and broadness of the corresponding diffraction spots in LEED. We explain this {\mbox{(10${\times}$10)}} order relative to graphene based on a grain boundary formation model for the Pb atoms at the graphene/SiC interface \cite{SCHADLICH2023}. Another similar model has been suggested recently, that explains it through a network of vacancy line defects forming Frenkel-Kontorova domains in the intercalated Pb layer \cite{VERA2024}. 
\begin{figure*}[t]
	\centering
    \includegraphics{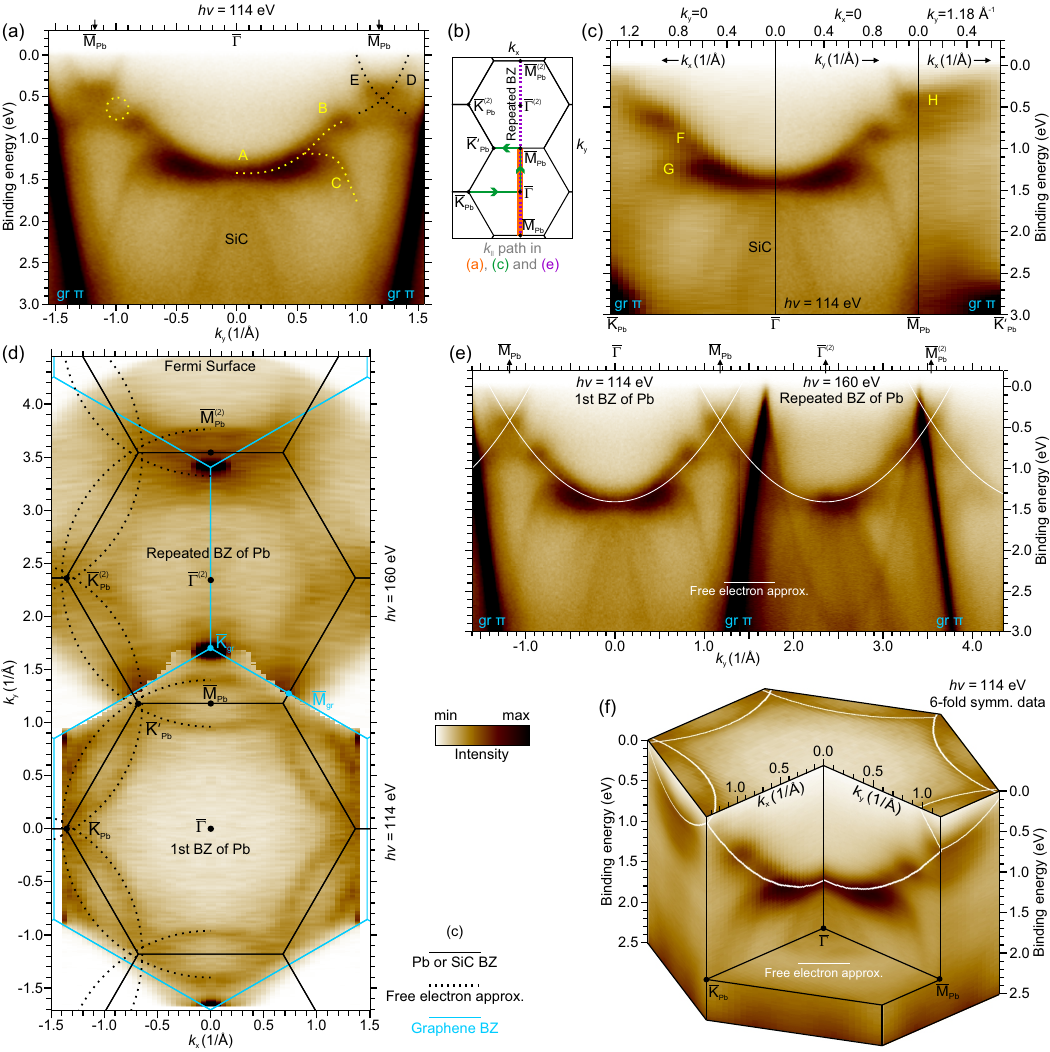}
	\caption{(a)  $E$-$k$ dispersion for Pb-QFMLG ($h\nu$ = $114$ eV) along the $\overline{\mathrm{M}\Gamma\mathrm{M}}$ direction of the Pb-BZ. (b) A geometrical representation of the first and repeated BZs of Pb, showing the momentum space paths considered in (a), (c) and (e), respectively. (c) $E$-$k$ dispersion along the $\overline{\mathrm{K}\Gamma\mathrm{MK'}}$ path of the Pb-BZ. (d) Fermi surface showing the first BZ ($h\nu$ = $114$ eV) and the repeated BZ ($h\nu$ = $160$ eV) of Pb, overlaid with the BZs of Pb (or SiC) and graphene in black and blue, respectively; the free electron approximation is overlaid as black dotted curves on the left half of the image. (e) $E$-$k$ dispersion along the $\overline{\mathrm{M}\Gamma\mathrm{M}}$ direction of Pb in its first and repeated BZs; free electron parabolas shown in white. (f) A 3D overview of the interlayer Pb bands ($h\nu$ = $114$ eV); free electron approximation overlaid in white.}
	\label{Fig_2}
\end{figure*}

\begin{figure*}[t]
	\centering
    \includegraphics{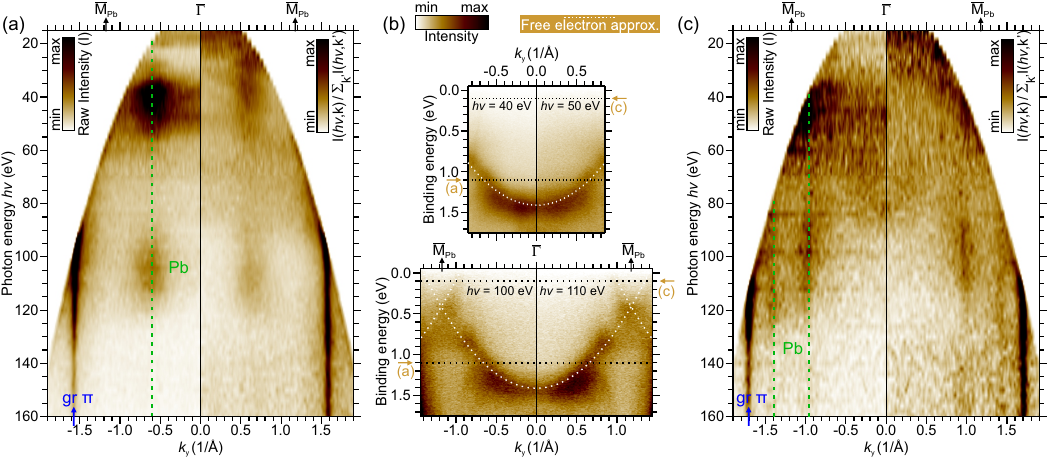}
	\caption{(a) CIS map along the $\overline{\mathrm{M}\Gamma\mathrm{M}}$ direction of Pb at $E = 1.1$ eV below the Fermi level. The Pb bands appear at |$\vec{k}_{||}$| $\approx 0.6$~{\AA$^{-1}$}, emphasized with a green dotted line in the left half; the blue arrow points at the graphene $\pi$-band around its $\overline{\mathrm{K}}$ point. The right half of the CIS map is normalized with respect to the momentum-integrated intensity. (b) $E$-$k$ cuts along the $\overline{\mathrm{M}\Gamma\mathrm{M}}$ direction of Pb at photon energies of $40$ and $50$ eV (top), $100$ and $110$ eV (bottom), respectively, overlaid with the free electron approximation. (c) CIS map at $E = 0.1$ eV below the Fermi level, showing two distinct Pb band features at about 0.96~{\AA$^{-1}$} and 1.4~{\AA$^{-1}$} around the $\overline{\mathrm{M}}$ point of Pb, emphasized with green dotted lines in the left half; the right half shows a normalized version similar to (a).}
	\label{Fig_3}
\end{figure*}

\subsection{Electronic band structure of interlayer Pb and its atomic arrangement relative to SiC}\label{sub2}
In our previous study, interlayer Pb bands were measured using He I photons ($h\nu$ = $21.22$ eV) \cite{MATTA2022}, where the prominent graphene $\pi$ bands and the SiC bulk bands overshadowed the Pb bands. It was therefore challenging to clearly resolve the Pb band structure away from the $\overline\Gamma$ point, especially closer to other high-symmetry points, as also observed with He II photons \cite{SCHOLZEL2024}. This highlights the need to investigate the Pb bands using different photon energies to enhance their photoemission cross section, as we have already demonstrated in \cite{SCHADLICH2023}. Reference \cite{VERA2024} also presents the interlayer Pb band structure measured at a higher photon energy of $110$ eV, but it covers only a limited momentum space and binding energy range, lacking a comprehensive overview. 
\\Based on photon energy dependent measurements (see Sec.~\ref{sub3}), we chose a photon energy of $114$ eV for detailed measurements of the interlayer Pb bands as shown in Fig.~\ref{Fig_2}. The $E$-$k$ dispersion along the $\overline{\mathrm{M}\Gamma\mathrm{M}}$ direction of the Pb-BZ \footnote{First and repeated BZ measurements for a Pb-QFMLG show that the surface BZ of interlayer Pb coincides with that of the SiC substrate, as discussed in Figs.~\ref{Fig_2}(d) and (e)} is shown in Fig.~\ref{Fig_2}(a). The SiC valence bands show a lower spectral intensity at $114$ eV when compared to our He I ARPES measurements in \cite{MATTA2022}. The overilluminated graphene $\pi$ bands (labeled as gr $\pi$) are visible around their respective $\overline{\mathrm{K}}$ and $\overline{\mathrm{K'}}$ points. The interlayer bands carry significant spectral weight with a local band minimum at $\overline\Gamma$ (yellow dotted curve 'A', as guide to the eye) and an electron-like upward dispersion towards the $\overline{\mathrm{M}}$ point of Pb. The latter splits into an upward and a downward branch around an in-plane momentum |$\vec{k}_{||}$| $\approx 0.6$~{\AA$^{-1}$}, marked as yellow dotted curves 'B' and 'C', respectively in Fig.~\ref{Fig_2}(a). Closer to the $\overline{\mathrm{M}}$ point, a band marked as 'D' (black dotted curve) disperses further upwards crossing the Fermi energy. Another band marked as 'E' (black dotted curve), crosses the band 'D' as well as the Fermi energy and seems to be a symmetric counterpart of band 'D' with respect to the $\overline{\mathrm{M}}$ point. Note that there is a suppression of spectral intensity around a binding energy range of $0.5$ to $0.8$ eV at |$\vec{k}_{||}$| $\approx 1.0~\AA^{-1}$ [yellow dotted circle in Fig.~\ref{Fig_2}(a)], which might be related to a change in the orbital character or a band gap. Fig.~\ref{Fig_2}(b) shows the momentum space paths considered in Figs.~\ref{Fig_2}(a), (c) and (e), respectively. 
Fig.~\ref{Fig_2}(c) displays the $\overline{\mathrm{K}\Gamma\mathrm{MK'}}$ path of the Pb-BZ. Similar to the $\overline{\Gamma\mathrm{M}}$ direction, the interlayer bands also show an upward electron-like dispersion along the $\overline{\Gamma\mathrm{K}}$ direction, again combined with a split into two branches (upward 'F' and downward 'G') at |$\vec{k}_{||}$| $\approx 0.6$~{\AA$^{-1}$}. Furthermore, the upward band 'F' along $\overline{\Gamma\mathrm{K}}$ shows reduced intensity from a binding energy of $\approx 0.5$ eV up to the Fermi energy. Along the $\overline{\mathrm{MK'}}$ direction, the interlayer band shows a relatively flat electron-like dispersion, marked as 'H', also having a gradual intensity suppression towards the $\overline{\mathrm{K'}}$ point.  
\\The Fermi surface of Pb-QFMLG integrated over a width of $\approx 60$ meV is shown in Fig.~\ref{Fig_2}(d) covering the first and repeated BZs of Pb \footnote{Here, the ARPES data obtained in half of the Fermi surface has been mirrored with respect to the $\overline{\mathrm{M}\Gamma\mathrm{M}}$ direction of Pb ($\overline{\mathrm{K}\Gamma\mathrm{K'}}$ direction of graphene)}. The Pb bands appear as nearly circular contours and can be fitted with a free electron approximation considering a Fermi momentum |$\vec{k}_F$| $\approx 1.4$~{\AA$^{-1}$} and an effective electron mass ($m^*$) of 5.3 times the rest electron mass ($m_e$), as shown by the black dotted circles. These circular contours repeat with the SiC lattice periodicity, hence demonstrating a {\mbox{(1${\times}$1)}} order of the electronic structure of interlayer Pb relative to SiC \cite{SCHADLICH2023}. These contours form an electron pocket around the $\overline{\mathrm{M}}$ point of Pb. Note that the part of this electron pocket falling in the first BZ corresponds to a repeated BZ contour and the counterpart that falls outside the first BZ boundary corresponds to the first BZ contour. According to Luttinger's theorem, the free electron approximation gives an electron density of about $6.8\times10^{14}$ cm$^{-2}$ for interlayer Pb \cite{LUTTINGER1960,otherLUTTINGER1960}. Furthermore, we present the $E$-$k$ dispersion along the $\overline{\mathrm{M}\Gamma\mathrm{M}}$ direction of Pb in its first and repeated BZs, overlaid with the free electron parabolas, see Fig.~\ref{Fig_2}(e). The quasi-free electron-like dispersion of the Pb bands indicates that the Pb-QFMLG heterostack exhibits Van der Waals interactions. Note that the measurements in the repeated BZ of Pb are taken at a photon energy of $160$ eV instead of $114$ eV. This was necessary in order to maximize the photoemission cross section for Pb in its repeated BZ, which significantly varies compared to the first BZ due to a different measurement geometry involving a tilt of about $26^{\circ}$ with respect to the normal emission orientation of the sample. The Fermi surface in the repeated BZ is shown separately in supplementary Fig. S1 \change{\cite{supplement}}, revealing $\pi$-band replicas around the $\overline{\mathrm{K}}$ point of graphene, corresponding to a {\mbox{(10${\times}$10)}} periodicity relative to the graphene unit cell, consistent with the LEED and STM results [Fig.~\ref{Fig_1}]. Furthermore, Fig. S2 in the supplementary information  \change{\cite{supplement}} shows the fitted Pb band dispersion along its $\overline{\mathrm{M}\Gamma\mathrm{M}}$ and $\overline{\mathrm{K}\Gamma\mathrm{K'}}$ directions, demonstrating a different band curvature along the respective high symmetry directions, evidently deviating from a free electron-like parabola. This indicates an anisotropic effective electron mass corresponding to the Pb bands, which is expected to have implications on charge transport measurements in the interlayer Pb.
\\Finally, a three-dimensional (3D) overview of the Pb band structure is shown in Fig.~\ref{Fig_2}(f) along with the free electron approximation. This is a six-fold rotationally symmetrized ARPES dataset, a procedure that enhances the signal-to-noise ratio for any weak band features, keeping the band structure information intact for a 2D hexagonal lattice arrangement, as in the case of Pb. This symmetrization will also be referred to in further discussions in this work. The band splittings, potential gaps and other renormalizations in the Pb band structure, shown in the detailed data evaluation, are beyond the scope of a simple free electron model. Thus, an advanced theoretical approach, such as DFT, is necessary to better understand the interlayer band structure, as discussed later in Sec.~\ref{sub4}. 

\begin{figure*}[t]
	\centering
    \includegraphics{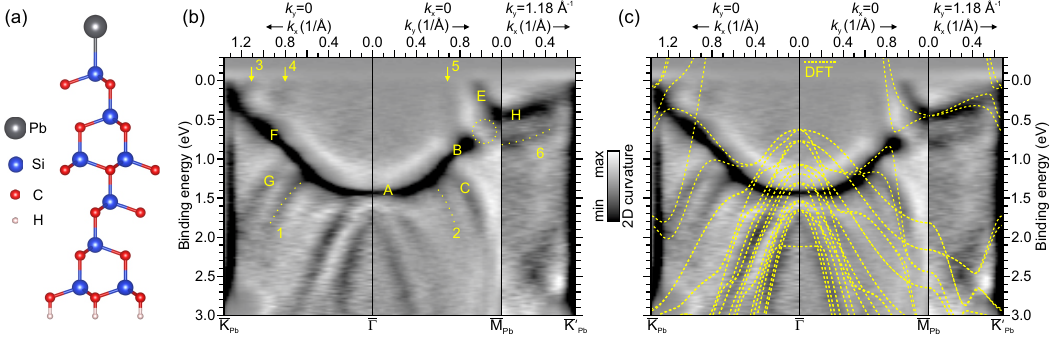}
	\caption{(a) Side view of atomic arrangement of a Pb {\mbox{(1${\times}$1)}} monolayer on 6H-SiC, considered for the DFT calculations displayed in (c). (b) 2D curvature of $E$-$k$ dispersion along the $\overline{\mathrm{K}\Gamma\mathrm{MK'}}$ path of the Pb-BZ ($h\nu$ = $114$ eV). (c) The same dataset as (b) overlaid with DFT calculations (yellow dotted curves). \change{Note that the minima in the 2D curvature (black color) correspond to regions of high spectral intensity \cite{ZHANG2011}.}}
	\label{Fig_4}
\end{figure*} 

\subsection{Photoemission cross section and two-dimensionality of interlayer Pb}\label{sub3}
In order to maximize the spectral intensity of the Pb bands, we scanned a photon energy range of $15$ to $160$ eV. Figs.~\ref{Fig_3}(a) and (c) display constant initial state (CIS) maps displaying the band dispersion relative to photon energy at a specific binding energy ($E$), along the $\overline{\mathrm{M}\Gamma\mathrm{M}}$ direction of the Pb-BZ. The left half of these CIS maps presents the raw data, while the right half shows the spectral intensity normalized with respect to the momentum-integrated intensity. The CIS map in Fig.~\ref{Fig_3}(a) is obtained at $E = 1.1$ eV, where the interlayer Pb bands are visible at |$\vec{k}_{||}$| $\approx 0.6~\AA^{-1}$ (green dotted line), demonstrating two distinct photon energy windows with a higher photoemission cross section for Pb, approximately $35$--$55$ eV and $95$--$115$ eV, respectively. Around the $35$--$55$ eV window, it is not possible to cover the Pb band dispersion up to the Fermi energy in a single deflection mode ARPES map. Additionally, the graphene bands have a higher photoemission cross section in this energy range, which further limits the visibility of the Pb bands. Therefore, for detailed Pb band measurements, we chose a photon energy of $114$ eV providing high spectral intensity, better visibility of the Pb bands relative to graphene, and a broad coverage of the momentum space in a single ARPES measurement. The right half of the CIS map highlights the non-dispersive nature of the Pb bands with respect to photon energy (effectively, with respect to the out of plane momentum, $k_z$), demonstrating the 2D character of the intercalated Pb layer. 
The graphene $\pi$ bands appear at |$\vec{k}_{||}$| $\approx 1.57~\AA^{-1}$, marked with a blue arrow \footnote{The $\pi$ bands do not appear at |$\vec{k}_{||}$| = $1.702~\AA^{-1}$ (momentum value corresponding to the $\overline{\mathrm{K}}$ point of graphene along its $\overline{\Gamma\mathrm{K}}$ direction) because the CIS map is not taken at the Fermi energy but instead at a binding energy of $1.1$ eV}. Fig.~\ref{Fig_3}(b) displays the $E$-$k$ dispersion of Pb along its $\overline{\mathrm{M}\Gamma\mathrm{M}}$ direction at different photon energies, illustrating its independence of the latter. Another CIS map at $E = 0.1$ eV [see Fig.~\ref{Fig_3}(c)] shows two distinguishable band features around the $\overline{\mathrm{M}}$ point of Pb (green dotted lines), corresponding to the first (|$\vec{k}_{||}$| $\approx 1.4~\AA^{-1}$) and the repeated BZ (|$\vec{k}_{||}$| $\approx 0.96~\AA^{-1}$) contours in the Fermi surface, as discussed in Fig.~\ref{Fig_2}(d). These bands close to the Fermi level also confirm a higher photoemission cross section for Pb around the photon energy windows discussed above. 

\begin{figure*}[t]
	\centering
    \includegraphics{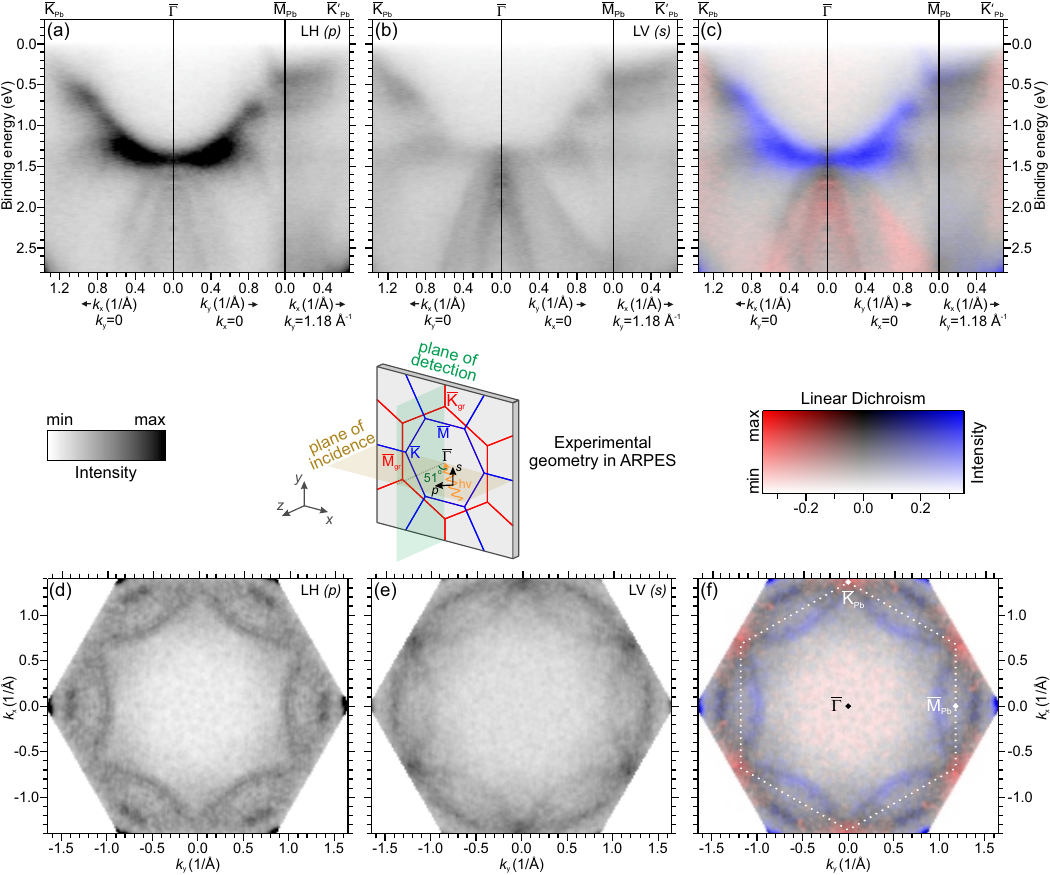}
	\caption{Influence of light polarization on the band intensities of interlayer Pb ($h\nu$ = $114$ eV): (a), (b) $E$-$k$ dispersion for the interlayer Pb bands measured along the $\overline{\mathrm{K}\Gamma\mathrm{MK'}}$ path of the Pb-BZ, using (a) linear horizontal (\textit{p}) and (b) vertical (\textit{s}) light polarizations. (c) Linear dichroism along the $\overline{\mathrm{K}\Gamma\mathrm{MK'}}$ path of Pb. (d), (e) CESs obtained at $E = 0.05$ eV, using (d) linear horizontal (\textit{p}) and (e) vertical (\textit{s}) light polarizations. (f) Linear dichroism obtained for the CES at $E = 0.05$ eV. Note that the spectral intensity in (d), (e) and (f) is scaled up by a factor of 6 compared to (a), (b) and (c), respectively.}
	\label{Fig_5}
\end{figure*}

\subsection{Comparison of interlayer Pb bands with DFT calculations} \label{sub4}
As discussed in Sec.~\ref{sub2}, the free electron approximation is not sufficient to explain the renormalizations observed for the Pb bands. Therefore, considering the {\mbox{(1${\times}$1)}} order of the interlayer Pb relative to SiC, we compare our ARPES data with DFT calculations for a {\mbox{(1${\times}$1)}} Pb monolayer on SiC; the respective atomic arrangement is displayed in Fig.~\ref{Fig_4}(a). These calculations exclude the graphene layer on top and align well with references \cite{VERA2024} and \cite{SCHOLZEL2024}. For a better signal-to-noise ratio for the interlayer bands, a 6-fold symmetrized ARPES dataset is used. To further enhance the data visualization, we use the 2D curvature which is a better alternative to a second derivative, with the advantages of improved tracking of extrema and reduced broadness of dispersive features \cite{ZHANG2011}. 
\\Fig.~\ref{Fig_4}(b) displays the 2D curvature of the interlayer Pb bands along the $\overline{\mathrm{K}\Gamma\mathrm{MK'}}$ path of the Pb-BZ, revealing additional downward band branches after the splitting of the electron-like Pb band at |$\vec{k}_{||}$| $\approx 0.6~\AA^{-1}$; marked as yellow dotted curves '1' and '2' towards the $\overline{\Gamma\mathrm{K}}$ and $\overline{\Gamma\mathrm{M}}$ directions, respectively. Apart from that, some weak spectral features crossing the Fermi level are revealed, marked by yellow arrows as '3', '4' and '5', respectively.
 Fig.~\ref{Fig_4}(c) shows the same ARPES dataset overlaid with the DFT calculations\change{, solely shown in Fig. S3 (b) \cite{supplement}}, which closely capture the splitting of the electron-like interlayer band as well as the anisotropic band curvature along the $\overline{\Gamma\mathrm{M}}$ and $\overline{\Gamma\mathrm{K}}$ directions of Pb. DFT shows a band gap of $\approx 180$ meV around $E \approx 0.75$ eV and |$\vec{k}_{||}$| $\approx 0.8~\AA^{-1}$ along $\overline{\Gamma\mathrm{M}}$, which \change{qualitatively aligns} with the intensity suppression observed experimentally, marked by a yellow dotted circle in Fig.~\ref{Fig_4}(b) [also discussed in Fig.~\ref{Fig_2}]. Moreover, DFT shows two band branches crossing the Fermi energy along $\overline{\Gamma\mathrm{M}}$, which is also qualitatively in line with our experimental observations. Along $\overline{\Gamma\mathrm{K}}$, DFT predicts a $\approx 25$ meV band gap around $E \approx 0.5$ eV and |$\vec{k}_{||}$| $\approx 0.9~\AA^{-1}$, which is not obvious in the experimental data. However, there is a clear intensity suppression of the band dispersion along $\overline{\Gamma\mathrm{K}}$ at $E \approx 0.5$ eV, as also mentioned in Sec.~\ref{sub2}. Around the $\overline{\mathrm{K}}$ point, DFT matches the features '3' and '4' crossing the Fermi energy. \change{Along $\overline{\mathrm{MK'}}$, a relatively flat upward dispersion ('H') at $E \approx 0.4$ eV is observed experimentally, while DFT predicts a similar feature as split bands. The absence of splitting may simply be due to the broadness of feature ‘H’, making it difficult to resolve. Additionally, a similarly dispersing band feature appears at $E \approx 0.7$ eV, highlighted by the yellow dotted curve '6' in Fig.~\ref{Fig_4}(b), which is not captured by theory. This discrepancy likely stems from the structural model, which omits the graphene layer and the {\mbox{(10${\times}$10)}} Pb superstructure relative to the graphene unit cell. Moreover, the raw spectral intensity along $\overline{\mathrm{MK'}}$ in Fig. 2(c) appears as a single broad feature ‘H’. Since raw data generally holds richer information \cite{ZHANG2011}, the apparent separation into two features in the 2D curvature might be misleading, suggesting that feature ‘6’ could actually originate from ‘H’ itself. However, the broadness of band 'H' in the raw data also contrasts with the theory, potentially due to structural approximations in the latter.} 
\change{Furthermore, DFT underestimates the binding energy of the SiC valence band maximum. ARPES measurements ($h\nu = 26$ eV) identify it at $E \approx 1.25$ eV (Fig. S4 \cite{supplement}), which is $\approx 600$ meV higher than what DFT predicts. This discrepancy arises from the sensitivity of the DFT calculations to the Pb-Si distance, where the ideal {\mbox{(1${\times}$1)}} Pb monolayer and the absence of the graphene layer do not seem to accurately capture the interlayer charge transfer between Pb and SiC. Nonetheless, our DFT calculations qualitatively capture the key features of the experimentally observed Pb band structure, which is the primary focus of this work.} 

\subsection{Light polarization dependence of interlayer band intensities: Orbital character}
In the ARPES measurements discussed so far, we have used linear horizontal light polarization, observing various intensity suppressions in the Pb bands, especially near the Fermi energy. These suppressions could be associated with a change in the orbital character of the bands. To gain a deeper qualitative understanding of the orbital character of different band features, we conducted light polarization dependent ARPES measurements, using linear horizontal (LH) and linear vertical (LV) light polarizations. Considering our ARPES experimental geometry, see inset in Fig.~\ref{Fig_5}, LH polarization is parallel to the plane of light incidence and hence corresponds to \textit{p}-polarized light; LV polarization is perpendicular to the plane of light incidence, hence \textit{s}-polarized. Fig.~\ref{Fig_5} displays six-fold symmetrized ARPES datasets obtained at a photon energy of $114$ eV. The band structure of Pb along the $\overline{\mathrm{K}\Gamma\mathrm{MK'}}$ path is shown in Figs.~\ref{Fig_5}(a) and (b), measured with LH and LV light polarizations, respectively. Fig.~\ref{Fig_5}(c) shows the linear dichroism along the same path, i.e., the intensity difference between the two light polarizations normalized to their sum. To this end, a 2D colorscale is used where the light polarization asymmetry is encoded along the horizontal axis and the intensity along the vertical axis. Maximum intensity of blue and red corresponds to most dominant spectral weights from LH and LV light polarizations, respectively. Note that a multiplicative factor was applied to normalize the spectral intensities in these measurements, compensating for differences in total flux between the two polarizations.
\\Based on the geometry of our ARPES experiments, LV polarization primarily probes the in-plane atomic orbitals of a sample, while LH polarization probes a mix of both in-plane and out-of-plane atomic orbitals. This is because the light beam does not probe the sample surface at grazing incidence but at an angle of $51^{\circ}$ with respect to the sample normal as shown in the central inset of Fig.~\ref{Fig_5}. As a result, Fig.~\ref{Fig_5}(c) suggests that the electron-like band around $\overline\Gamma$ is predominantly due to the out-of-plane atomic orbitals. Along the $\overline{\Gamma\mathrm{M}}$ direction of Pb, the out-of-plane orbital character remains dominant, though it gradually weakens towards the $\overline{\mathrm{M}}$ point, suggesting contributions from in-plane atomic orbitals. Higher statistics ARPES measurements along the $\overline{\mathrm{M}\Gamma\mathrm{M}}$ direction of Pb, using both polarizations, are shown in Fig. S5 \change{\cite{supplement}}. Along the $\overline{\mathrm{MK'}}$ direction, the color gradually shifts from blue to grey, suggesting a gradual mixing of in-plane and out-of-plane atomic orbitals. This transition continues, with the orbital character predominantly switching to in-plane around the $\overline{\mathrm{K'}}$ point. Similarly, along the $\overline{\Gamma\mathrm{K}}$ direction, the electron-like band exhibits a strong out-of-plane character around $\overline{\Gamma}$, which gradually decreases, transitioning into a mix of in-plane and out-of-plane orbital characters, eventually becoming predominantly in-plane around the $\overline{\mathrm{K}}$ point. Figs.~\ref{Fig_5}(d) and (e) show constant energy surfaces (CESs) at $E = 0.05$ eV below the Fermi level, integrated over a width of $40$ meV, obtained with LH and LV light polarizations, respectively. Fig.~\ref{Fig_5}(f) illustrates the corresponding linear dichroism, clearly demonstrating that the orbital character is predominantly out-of-plane across the entire BZ, except near the $\overline{\mathrm{K}}$ point of Pb. 
\\According to our DFT calculations, a {\mbox{(1${\times}$1)}} Pb monolayer on SiC exhibits a dominant $p$-orbital character. DFT calculations (without SOC) showing the corresponding in-plane ($p_{x}$ and $p_{y}$) and out-of-plane ($p_{z}$) atomic orbital weights, are provided in supplementary information [Figs. S3 (c) and (d)]  \change{\cite{supplement}}. The calculations reveal a $p_{z}$ orbital character for most of the Pb bands, while band features near the $\overline{\mathrm{K}}$ point of Pb display contributions from $p_{x}$ and $p_{y}$ orbitals, which closely aligns with our experimental findings. \change{Furthermore, these calculations suggest that the electron-like Pb band around $\overline{\Gamma}$ hybridizes with the SiC bulk bands \cite{MATTA2022}, which might explain the flattening of the dispersion around $\overline{\Gamma}$, as seen in Fig. S2 \cite{supplement}. Moreover, the} dominance of $p_{z}$ orbital character in Pb bands is noteworthy, as it suggests potential strong interactions between the intercalated Pb atoms and the $\pi$ orbitals of graphene, possibly contributing to proximity effects in graphene. Notably, DFT also predicts a $p_{z}$ band just below the Fermi energy near the $\overline{\mathrm{K}}$ point of Pb \change{[Fig. S3 (d)] \cite{supplement}}, where our experimental data shows only a minor contribution from LH polarization. \change{Fig. S3 (b) \cite{supplement} shows that the spectral weight of this band effectively decreases below the Fermi energy when the SOC is included. Hence, it is a subtle discrepancy, that can be attributed to the reduced complexity of our structural model.}

\subsection{Charge transfer into the quasi-freestanding graphene layer}

Earlier ARPES measurements at room temperature have revealed the near charge-neutrality of Pb-QFMLG with a slight residual $p$-doping of less than $2\times10^{10}$ cm$^{-2}$ \cite{MATTA2022,SCHADLICH2023}. This level of doping is negligibly small when compared to other lightly or moderately doped epitaxial graphene systems \cite{RIEDL2009,EMTSEV2011,FORTI2020}. There has been some speculation about the charge transfer mechanisms responsible for this near charge-neutrality of Pb-QFMLG. Two potential mechanisms have been proposed; (i) Any charge transfer into graphene from the intercalated Pb atoms and the SiC substrate might cancel each other out. (ii) the intercalated Pb layer might screen the charge transfer from the substrate, while not transferring any charge itself to the graphene layer above \cite{MATTA2022,SCHADLICH2023}. Schölzel \textit{et al.} \cite{SCHOLZEL2024} propose that the second mechanism is at play, based on measurements of Pb-QFMLGs prepared on 6H- and 4H-SiC substrates. Here, we present a different approach to investigate the substrate's influence on the graphene doping. For a doped semiconductor, like the $n$-doped SiC in our case, the bulk dopants are expected to freeze out with decreasing temperatures. Therefore, we measured the graphene $\pi$ bands at different temperatures in order to obtain insight into a potential charge transfer from the substrate. Complete screening of the SiC substrate by the Pb layer would result in a largely temperature-independent doping of graphene. On the other hand, if a balance of charge transfers from Pb and SiC occurs, graphene is expected to become more $p$-doped as the temperature decreases, due to fewer electrons being released from the SiC substrate. It is well known that the $n$-dopants in SiC reduce the overall $p$-doping in H-intercalated epitaxial graphene when compared to a semi-insulating SiC substrate \cite{MAMMADOV2014}. 
\begin{figure}[t]
	\centering
    \includegraphics{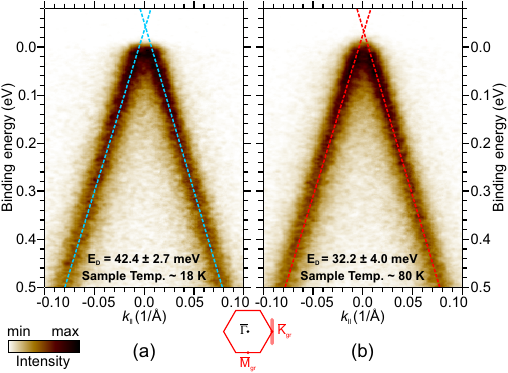}
	\caption{$E$-$k$ dispersion ($h\nu$ = 40 eV) at the $\overline{\mathrm{K}}$ point of graphene, perpendicular to its $\overline{\Gamma\mathrm{K}}$ direction at a sample temperature of about (a) $18$ K and (b) $80$ K, overlaid with blue and red colored dotted linear fits (obtained using MDC fits through the dispersion), respectively.}
	\label{Fig_6}
\end{figure}
\\Figs.~\ref{Fig_6}(a) and (b) show $E$-$k$ cuts for Pb-QFMLG (same sample under the same measurement conditions except sample temperature) obtained at the $\overline{\mathrm{K}}$ point of graphene, perpendicular to its $\overline{\Gamma\mathrm{K}}$ direction, measured with a photon energy of $40$ eV at sample temperatures of approximately $18$ K and $80$ K, respectively. From the momentum distribution curve (MDC) fitting of the Dirac cone dispersion, we find the Dirac point energy ($E_D$) to be $42.4\pm2.7$ meV at $18$ K, and $32.2\pm4.0$ meV at $80$ K, above the Fermi energy, corresponding to hole densities of $(1.2\pm0.2)\times10^{11}$ cm$^{-2}$ and $(7.0\pm1.7)\times10^{10}$ cm$^{-2}$, respectively. This corresponds to an increase in the $p$-doping of Pb-QFMLG by $(5.0\pm2.6)\times10^{10}$ cm$^{-2}$ ($\approx1.7$ times higher), when decreasing the sample temperature from $80$ K to $18$ K. This increase in $p$-doping can be attributed to the freezing out of $n$-type dopants of SiC, suggesting that the near charge-neutrality of Pb-QFMLG is a combined effect of charge transfers from the intercalant Pb and the SiC substrate. Therefore, this rules out the possibility of a complete screening of charge transfer from the SiC substrate by the interlayer Pb. For reference, a room temperature ARPES measurement of the same Pb-QFMLG sample is provided in supplementary information \change{\cite{supplement}}, where $E_D$ is found to be $13.5\pm4.6$ meV above the Fermi level, corresponding to a hole density of $(1.2\pm0.8)\times10^{10}$ cm$^{-2}$, as shown in Fig. S6 \change{\cite{supplement}}. By using the same sample for these temperature dependent measurements, we can rule out any uncertainities in the carrier density of graphene associated with the initial quality of ZLG, the degree of intercalation or the formation of distinct Pb phases at the graphene/SiC interface.

\section{CONCLUSION AND OUTLOOK}
Synchrotron-based ARPES measurements, using different photon energies and light polarizations, reveal several key characteristics of the interlayer Pb band structure. These interlayer bands can be fitted with a free electron approximation to a first order, revealing the Van der Waals nature of the interlayer forces in a Pb-QFMLG heterostack. CIS mapping identifies two distinct photon energy windows where the Pb bands exhibit a higher photoemission cross section, amongst which the higher photon energies allow measuring the Pb bands with high spectral intensity over the entire BZ in a single ARPES deflection map. Repeated BZ measurements demonstrate a {\mbox{(1${\times}$1)}} order of Pb relative to the SiC substrate. Additionally, the {\mbox{(10${\times}$10)}} periodicity with respect to the graphene unit cell, observed in LEED and STM, is also corroborated by ARPES through the observation of graphene replicas.
\\DFT calculations for a {\mbox{(1${\times}$1)}} Pb monolayer on SiC successfully capture features such as band splittings, intensity suppressions or gaps, and anisotropic band curvature in the experimentally observed interlayer band structure. ARPES measurements with different light polarizations together with DFT calculations uncover a predominant $p_{z}$ orbital character for the Pb bands, except around the $\overline{\mathrm{K}}$ point of Pb. This is particularly relevant for understanding potential proximity effects in graphene induced by the interlayer Pb. 
\\Furthermore, we observe a doping change in the graphene $\pi$ bands with temperature, which we attribute to the freeze out of bulk $n$-dopants in SiC. At lower temperatures, fewer $n$-dopants can be released from the SiC substrate, resulting in an increased $p$-doping of graphene. This effect signifies the combined role of both the SiC substrate and the intercalated Pb in the near charge-neutrality of Pb-QFMLG at room temperature. 
\\Our findings provide a solid foundation for further detailed studies of the interlayer Pb band structure. For instance, Yang \textit{et al.} \cite{YANG2022} recently predicted a non-trivial ‘antivortex’ spin texture in a monolayer Pb on SiC, which could have intriguing implications for spin-transport phenomena. Such a prediction could be experimentally verified by conducting spin-resolved ARPES measurements on a Pb-QFMLG system in the future. 


\begin{acknowledgments}
This work was supported by the Deutsche Forschungsgemeinschaft (DFG, German Research Foundation) within the FLAG-ERA framework through Project Sta315/9-1 and within the Research Unit FOR5242 through Projects Ku4228/1-1 and Sta315/13-1. MAX IV Laboratory is acknowledged for time on beamline BLOCH under proposal 20210067, 20220474 and 20221217. We are also thankful to the entire beamline staff at BLOCH for their great support. Research conducted at MAX IV, a Swedish national user facility, is supported by the Swedish Research council under contract 2018-07152, the Swedish Governmental Agency for Innovation Systems under contract 2018-04969, and Formas under contract 2019-02496.
\end{acknowledgments}

\newpage
\onecolumngrid
\begin{center}
\textbf{\large\emph{Supplementary Information}}\\[0.1cm]
\textbf{\large Pb-intercalated epitaxial graphene on SiC: Full insight into band structure and orbital character of interlayer Pb, and charge transfer into graphene}
  \end{center}
  \setcounter{figure}{0}
  \renewcommand{\thefigure}{S\arabic{figure}}
  \vspace{\fill}
 \begin{figure*}[h]
	\centering
    \includegraphics{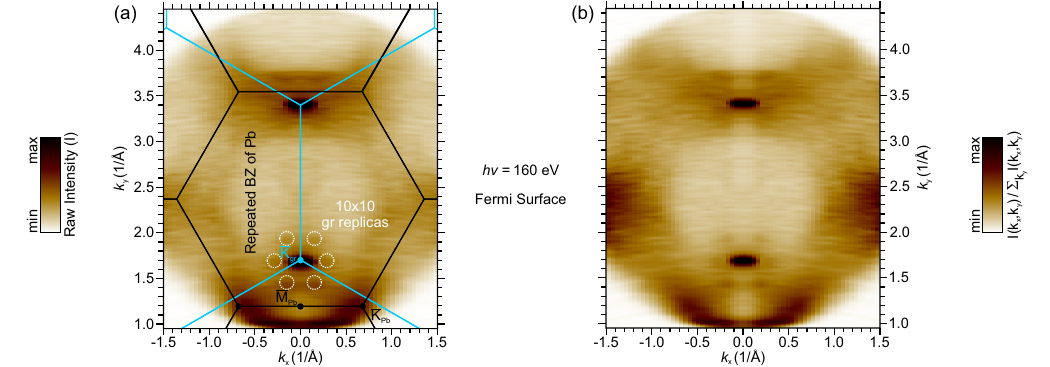}
	\caption{Fermi Surface for Pb-QFMLG obtained at $h\nu$ = $160$ eV, covering the repeated BZ of Pb [similar measurement as shown in Fig. 2(d) in the main text]; (a) Raw data with the Pb and graphene BZs overlaid in black and blue, respectively. Additionally, there is spectral intensity from replicated graphene Dirac cones around the $\overline{\mathrm{K}}$ point of graphene, marked in white circles. The average distance of the replicas from the $\overline{\mathrm{K}}$ point is $\approx 0.29~\AA^{-1}$, corresponding to a periodicity of about {\mbox{(10${\times}$10)}} relative to the graphene unit cell. (b) Same data as (a) with the spectral intensity normalized to the $k_y$-integrated intensity, visually enhancing the presence of the {\mbox{(10${\times}$10)}} graphene replicas.}
	\label{S1}
\end{figure*}

\begin{figure*}[h]
	\centering
    \includegraphics{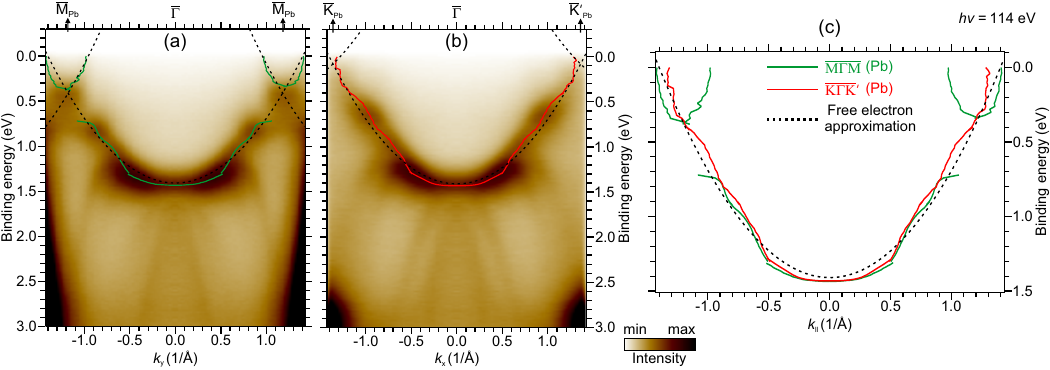}
	\caption{(a), (b) Interlayer Pb band dispersion along the $\overline{\mathrm{M}\Gamma\mathrm{M}}$ and $\overline{\mathrm{K}\Gamma\mathrm{K'}}$ directions, overlaid with combined MDC and EDC fits in green and red, respectively. Black dotted curves represent the free electron approximation, showing deviations from the true band dispersion. (c) The dispersion fits and the free electron approximation displayed together, revealing a different curvature of the electron-like Pb band along its $\overline{\Gamma\mathrm{M}}$ and $\overline{\Gamma\mathrm{K'}}$ directions, indicating an anisotropic effective electron mass.}
	\label{S2}
\end{figure*}

\begin{figure*}[h]
	\centering
    \includegraphics{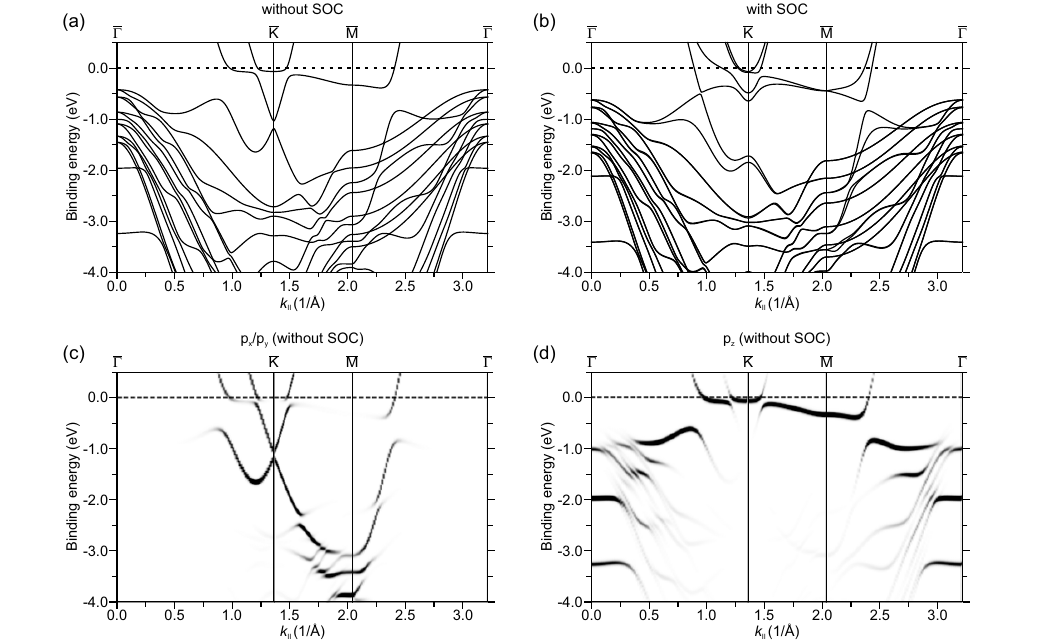}
	\caption{DFT calculations for a {\mbox{(1${\times}$1)}} Pb monolayer on SiC shown without and with spin-orbit coupling in (a) and (b), respectively. (c), (d) $p$-orbital projected Pb bands without spin-orbit coupling; the in-plane $p_x$ and $p_y$ orbital components are shown in (c), and the out-of-plane $p_z$ component in (d).}
	\label{S3}
\end{figure*} 

\begin{figure*}[h]
	\centering
    \includegraphics{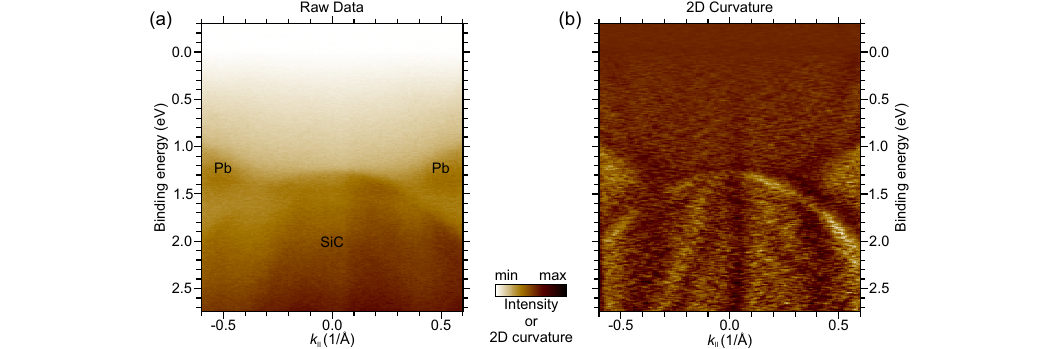}
	\caption{$E$-$k$ dispersion of Pb-QFMLG along the $\overline{\mathrm{M}\Gamma\mathrm{M}}$ direction of Pb at $h\nu$ = $26$ eV, with the raw data shown in (a) and the 2D curvature in (b) illustrating the valence band maximum of SiC at a binding energy of $\approx 1.25$ eV.}
	\label{S4}
\end{figure*}

\begin{figure*}[h]
	\centering
    \includegraphics{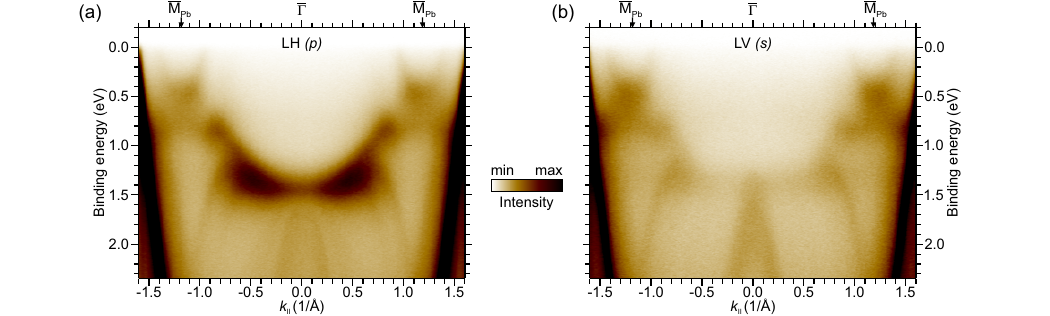}
	\caption{Interlayer Pb bands along the $\overline{\mathrm{M}\Gamma\mathrm{M}}$ direction of the Pb-BZ, obtained at $h\nu$ = $110$ eV with LH (\textit{p}) polarization in (a) and LV (\textit{s}) polarization in (b), respectively. The band intensity of the free-electron like parabola significantly drops around $\overline{\Gamma}$ when switching from LH (\textit{p}) to LV (\textit{s}) light polarization. This clearly demonstrates that the interlayer Pb bands around $\overline{\Gamma}$ exhibit a predominant out-of-plane character. Moreover, the band dispersion around the $\overline{\mathrm{M}}$ point of Pb has a considerable spectral weight in both measurements, suggesting a mix of both in-plane and out-of-plane orbital characters, as also discussed in the Sec. III E of the main text.}
	\label{S5}
\end{figure*}

\begin{figure*}[h]
	\centering
    \includegraphics{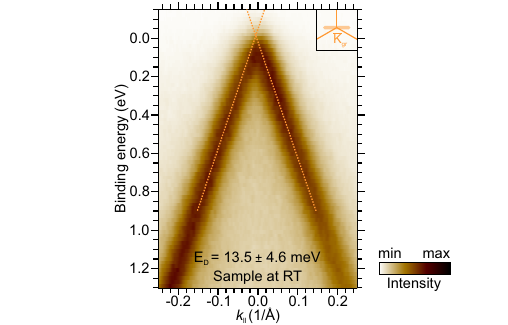}
	\caption{$E$-$k$ dispersion of Pb-QFMLG at the $\overline{\mathrm{K}}$ point of graphene, perpendicular to the $\overline{\Gamma\mathrm{K}}$ direction of graphene-BZ, obtained at room temperature using a home-lab based SPECS PHOIBOS 150 hemispherical analyzer with He II photons ($40.8$ eV). Based on MDC fits, the Dirac point energy turns out to be $13.5\pm4.6$ meV above the Fermi energy, corresponding to a $p$-type carrier density of $(1.2\pm0.8)\times10^{10}$ cm$^{-2}$. This is a reference measurement for the temperature dependent Dirac cone measurements shown in Fig. 6 of the main text, using the same Pb-QFMLG sample.}
	\label{S6}
\end{figure*}

\end{document}